\documentclass[prl,twocolumn,floatfix,showpacs]{revtex4}
\usepackage{graphicx}
\usepackage{amsmath}
\usepackage{amssymb}
\usepackage{mciteplus}
\usepackage{color}
\usepackage{nccmath}

\begin{document}
\title{Majorana-Klein hybridization in 
topological superconductor junctions}
\author{B. B\'eri}
\affiliation{TCM Group, Cavendish Laboratory, University of Cambridge, J.~J.~Thomson Ave., Cambridge CB3~0HE, UK}
\date{December 2012}
\begin{abstract}
We present a powerful and general approach to describe the coupling of Majorana fermions to external leads, of interacting or non-interacting electrons. Our picture has the Klein factors of bosonization appearing as extra Majoranas hybridizing with the physical ones. We demonstrate the power of this approach 
by solving a highly nontrivial SO($M$) Kondo problem arising in topological superconductors with $M$ Majorana-lead couplings, allowing for arbitrary $M$ and  for conduction electron interactions. We find that these topological Kondo problems give rise to robust non-Fermi liquid behavior, even for Fermi liquid leads, and to a quantum phase transition between insulating and Kondo regimes when the leads form Luttinger liquids.
In particular, for $M=4$ we find a long sought-after stable realization of the two-channel Kondo fixed point. 
\end{abstract}
\pacs{73.23.-b,74.78.Na,72.10.Fk,03.67.Lx}
\maketitle

One of the most influential recent discoveries in condensed matter physics is 
that topological phases supporting Majorana fermions can be engineered in heterostructures based on s-wave superconductors and materials with strong spin-orbit coupling\cite{FuKane08,*sau2010generic,*alicea2010majorana,*oreg2010helical}. 
This breakthrough development, transforming so-far elusive ideas\cite{kitaev2003fault} for Majorana based fault-tolerant quantum computers into a potentially feasible perspective, drives immense theoretical and experimental activities (see Ref.~\onlinecite{BeeMajrev,*AliMajrev} for reviews).

Majorana fermions are localized, robust, zero-energy excitations.
Testing the zero bias anomalies\cite{LawMaj,*Flencond,*Saucond,*wimmer2011quantum,Fid12} arising from electrons tunneling onto them forms one of the main experimental directions with promising results so far\cite{Mourik25052012,*deng2012observation,*das2012zero}. Transport can also inform on the Majoranas' quantum computational potential, if it indicates that the nonlocal ``topological qubits", formed of pairs of them obey quantum dynamics\cite{BeriTK}.

\begin{figure}
\includegraphics[width=0.8\columnwidth,clip=true,trim= 0 0 90 0]{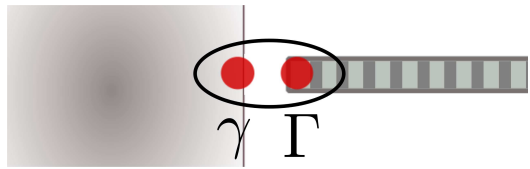}
\caption{A Majorana mode $\gamma$ on a superconductor heterostructure 
coupled to a lead of conduction electrons. The Majorana $\Gamma$ arises from the conduction electrons  $\psi =\Gamma f$. It hybridizes with $\gamma$ while $f$ changes the charge density (shown with waves).}
\label{fig:pictures}
\end{figure}

In all transport problems, a key role is played by the coupling of Majorana fermions $\gamma$ to leads of conduction electrons $\psi$. Here we introduce a general picture of this coupling (Fig.~\ref{fig:pictures}), observing that $\psi=\Gamma f$ can be broken into Majorana ($\Gamma$) and charge density ($f$) parts using bosonization. Upon $\Gamma$ hybridizing with the physical Majorana, the charge sector couplings get organized in a simple structure. 
Rendering the charge sector transparent is a key virtue of our picture. It allows one to approach qualitatively new problems involving the interplay of Majorana fermions and electron interactions, be those between conduction electrons or due to the charging of the superconductor.

We demonstrate this by solving a highly nontrivial problem that includes both types of interactions. This is the topological Kondo effect\cite{BeriTK}, which, for $M$ Majorana-lead couplings implements a novel, SO($M$) Kondo problem. In Ref.~\onlinecite{BeriTK}, we could solve this for the simplest, SO($3$) $\sim$ SU($2$) case for noninteracting leads. Our picture lets us explore vastly more general settings, allowing for arbitary $M$ and interacting conduction electrons. We find a number of striking features. These include stable, non-Fermi liquid (NFL) behavior, even for Fermi liquid leads; a quantum phase transition between Kondo and insulating regimes due to the competition between the Kondo effect and the suppression of electron tunneling in Luttinger liquids; and a long sought-after, stable realization of the two-channel Kondo fixed point for $M=4$. Our picture delivers these results with ease through mapping the problem to a quantum Brownian motion (QBM) model.

We begin by setting up our Majorana-lead coupling picture more explicitly. We consider a superconducting structure with $M_\text{tot}$  localized Majoranas $\gamma_j$. Their quasiparticle operators obey\cite{BeeMajrev}
\begin{equation}
\gamma_j=\gamma_j^\dagger,\quad\{\gamma_j,\gamma_k\}=2\delta_{jk}.\label{eq:MajAC}
\end{equation}
Ordinary fermionic modes $c_{jk}\!\!=\!\!\frac{1}{2}(\gamma_j\!+\!i\gamma_k)$ arise from pairs of Majoranas; in a fermionic system $M_\text{tot}$ is thus even. 

We will work with half infinite $(x\geq 0)$, single-channel leads furnishing effectively spinless conduction electrons. This can be achieved in several  Majorana realisations\cite{FuKane08}, including the nanowire based setups of recent experiments\cite{Mourik25052012}. 
Bosonization\cite{Giabook}  transforms the Hamiltonian of lead $j$ into a quadratic problem,
\begin{equation}
H_0(\varphi_j,\theta_j)=\frac{\hbar u}{8\pi}\int dxK(\partial_{x}\theta_j)^{2}+K^{-1}(\partial_{x}\varphi_j)^{2},
\label{eq:HLutt}
\end{equation}
even if interactions are present. Here $\theta_j,\varphi_j$ are bosonic fields encoding the charge density $\rho_j=\frac{\partial_x\varphi_j}{2\pi}$. They obey
\begin{equation}
[\varphi(x),\theta(y)]=4\pi i\Theta(x-y),
\end{equation}
where $\Theta(x)$ is the Heaviside function,
and satisfy\cite{Nayak99,*Oshikawa06}  $\varphi(0)=(\partial_x\theta)(0)=0$ at the endpoint of the lead.
Electron interactions enter Eq.~\eqref{eq:HLutt} through the Luttinger parameter $K$ and through renormalizing the velocity $u$. We have $K<1$ ($K>1$) in the repulsive (attractive) and $K=1$ in the noninteracting regimes.

Working at energies much below the gap, $M$ leads couple to the superconductor through $M$ Majoranas: 
\begin{equation}
H_t=\sum_{j=1}^{M}t_{j}\gamma_{j}\psi_{j}e^{i\hat\chi/2}+\text{H.c.},\label{eq:emcoupl}\end{equation} 
where $t_j$ is the tunneling amplitude, $\psi_j$ annihilates electrons at the end of lead $j$, 
and the phase exponential $e^{\pm i\hat\chi/2}$ changes the number of electrons on the superconductor by $\pm 1$.

Bosonization fractionalizes the electron operator as $\psi_j(x)=\Gamma_j f_j(x)$, 
where\cite{Nayak99,*Oshikawa06} 
$f_j=\frac{i}{\sqrt{a}} e^{i\theta_j/2}$
at the endpoint, up to a numerical factor to be absorbed in $t_j$ and with being $a$ the short distance cutoff. The Klein factors $\Gamma_j$ are often omitted but for us they are crucial. In ensuring that $\psi_j$ $\psi_k$, $\gamma_l$ anticommute, they extend  
Eq.~\eqref{eq:MajAC}:
\begin{equation}
\Gamma_j=\Gamma_j^\dagger,\quad\{\Gamma_j,\Gamma_k\}=2\delta_{jk},\quad \quad\{\Gamma_j,\gamma_l\}=2\delta_{jl}.
\end{equation}
This allows one to view $\Gamma_j$ as additional 
Majorana fermions which can, for example, form ordinary fermion modes $d_j=\frac{1}{2}(\gamma_j+i\Gamma_j)$ with the physical Majoranas. It also means that the terms in Eq.~\eqref{eq:emcoupl}
\begin{equation}
H_t=\sum_{j=1}^{M}\frac{it_j}{\sqrt{a}}\left(\gamma_{j}\Gamma_j\right)\left( e^{i\theta_j/2} e^{i\hat\chi/2}\right)+\text{H.c.}\label{eq:emcoupl1}\end{equation} 
factorize into Majorana-Majorana and charge sector parts. This is our main observation.

Before applying this to the topological Kondo problem, we warm up with the setup of Ref.~\onlinecite{Fid12}: tunneling  between a superconductor and a  Luttinger liquid lead ($M=1$). Recovering the couplings obtained there is a good validity check of our picture, which also allows us to highlight some of its advantages. The superconductor is grounded in this setup, thus we can drop  $e^{i\hat\chi/2}$ for now. 
If the system is in the topological regime the coupling is through a single Majorana, which enters the problem only through $\gamma_1\Gamma_1$. This is related to the parity $d_1^\dagger d_1\!\!=\!\!\frac{1}{2}(1+i\gamma_1\Gamma_1)$ and acts as $\sigma_1$ in the two state space of a $c_{1j}$ fermion (in the $(-1)^{c_{1j}^\dagger c_{1j}}\!\!=\!\!\sigma_3$ basis). This immediately recovers the coupling of Ref.~\onlinecite{Fid12}, circumventing the need for the augmented Jordan-Wigner transformation employed there.

In the nontopological regime, there is another Majorana $\gamma_2$ which couples to the lead and $\gamma_1$. The phase of $c_{12}$ and $\psi$ can always be chosen so that the coupling is
\begin{equation}
H_t=it_1\gamma_{1}\Gamma_1\frac{\cos(\theta_1/2)}{\sqrt a}+it_1^\prime\gamma_2\Gamma_1\frac{\sin(\theta_1/2)}{\sqrt a} +i\delta \gamma_1\gamma_2
\end{equation} 
with real $t_1,t_1^\prime,\delta$. Rotating from $\gamma_1,\gamma_2,\Gamma_1$ to new Majoranas $\tilde\gamma_{1,2,3}$  decouples $\tilde\gamma_{3}$ to give
\begin{equation}
H_t=i\tilde{\gamma}_1\tilde{\gamma}_2\sqrt{\delta^2+\frac{t_1^{\prime 2}}{a}+\frac{t_1^2-t_1^{\prime 2}}{a}\cos^2(\theta_1/2)}.
\end{equation}
This immediately recovers two key results of Ref.~\onlinecite{Fid12}: 
expanding the root for $t_1^2\approx t^{\prime 2}_1$ (which holds in the nontopological case) we find that the leading perturbation, which is also the most relevant in the sense of the renormalization group (RG), is  Cooper pair tunneling $\propto \cos^2(\theta_1/2)$; and that for $t_1^2\gg t_1^{\prime 2},a\delta^2$ near the topological regime the $\delta$ and $t_1^\prime$ terms give the same perturbation.

We now move on to the topological Kondo effect\cite{BeriTK}. In this Kondo effect strong correlations emerge even for noninteracting leads from the interplay of the charging energy $E_c$ of the superconducting island (which is now connected to ground by a capacitor) hosting the Majoranas and the ground state degeneracy associated with them\cite{BeeMajrev}. This degeneracy can be understood in terms of the fermion modes $c_{jk}$: they have zero energy  (up to exponentially small corrections in the Majoranas' separation) which leads to a $2^{M_\text{tot}-1}$ fold degenerate ground state, where the $(-1)$ in the exponent is because $\sum c^\dagger_{jk}c_{jk}$ mod $2$ is tied to the electron number parity of the island.

The fermion modes $c_{jk}$ provide the topological qubits and the Kondo effect arises from coupling these to conduction electrons via the Majoranas. For low temperatures, voltages and weak coupling, $T,V,|t_j|\ll E_c$, the physics is dominated by virtual transitions connecting the lowest energy  charge state of the island to the neighboring ones with $\pm1$ extra electrons. 
These are captured by the effective Hamiltonian  
\begin{equation}
H_{\text{eff}}=\sum_{j\neq k=1}^M\lambda^{+}_{jk}\gamma_{j}\gamma_{k}\psi_{k}^{\dagger}\psi_{j}-\sum_{j=1}^M\lambda^{-}_{jj}\psi_{j}^{\dagger}\psi_{j},
\label{eq:heffgam}
\end{equation}
where $\lambda^\pm_{jk}\sim t_{j}t_{k}/E_c$, $\lambda^+_{jk}>0$.  Eq.~\eqref{eq:heffgam} is obtained by  a Schrieffer-Wolf transformation keeping the leading order terms in $t_j/E_c$. As shown in Refs.~\onlinecite{BeriTK}, 
the nontrivial lead-qubit couplings of the first term implement an SO($M$) Kondo problem (with a spinor ``impurity spin" through $\gamma_j\gamma_k$). Ref.~\onlinecite{BeriTK} solved  
Eq.~\eqref{eq:heffgam}
for SO($3$)$\sim$SU($2$), applying the Affleck-Ludwig conformal field theory method\cite{affleck1990current,*affleck1991kondo,*affleck1991critical}. This is however unsuited for interacting leads and becomes complicated for $M>3$. As we now show, our picture handles both challenges with ease.

Substituting $\psi_j=\Gamma_j f_j$ into Eq.~\eqref{eq:heffgam}, we again find that Majoranas enter through  $\gamma_j\Gamma_j$\cite{M4fn}. Importantly, $\gamma_j\Gamma_j$ commute with each other and thus can be diagonalized simultaneously, $\gamma_j\Gamma_j=\pm i$, where we can absorb the sign in $\theta_j$. We get 
\begin{equation}
H_{\text{eff}}=-\sum_{j\neq k}\lambda_{jk}^{+}\frac{e^{-i\theta_{k}/2}e^{i\theta_{j}/2}}{a}-\sum_{j}\frac{\lambda_{jj}^{-}}{2\pi}\partial_{x}\varphi_{j}.\label{eq:bosocoupling}
\end{equation}
Eq.~\eqref{eq:bosocoupling} is the central formula underlying all our subsequent 
analysis. We reduced the Kondo problem into 
one for the charge densities, eliminating the Majorana degrees of freedom through the $\gamma_j\Gamma_j$ hybridizations. 

For $\lambda^-\!=0$, Eq.~\eqref{eq:bosocoupling} is known to map to QBM 
which,  for $\lambda^+_{jk}\!=\!\lambda^+\!<\!0$, $K\!>\!1$, is related to 
the $M\!\geq\! 3$ channel Kondo model\cite{YiKane,*YiQBM,Nayak99,*Oshikawa06,TeoKaneQPC}.
Our SO($M$) problem is in the $\lambda^+\!>\!0$ sector, 
and we have $K\leq 1$ in its lead-Majorana implementation. 
Eq.~\eqref{eq:bosocoupling}, for $M=3, \lambda^+_{jk}=\lambda^+, \lambda^-_{jj}=0$, also appeared as an unphysical description of quantum wire tri-junctions, used for illustrating the dangers of omitting Klein factors\cite{Nayak99,*Oshikawa06}. For us Eq.~\eqref{eq:bosocoupling} arises precisely from using Klein factors correctly.

\begin{figure}
\includegraphics[width=\columnwidth,clip=true,trim= 0 0 0 0]{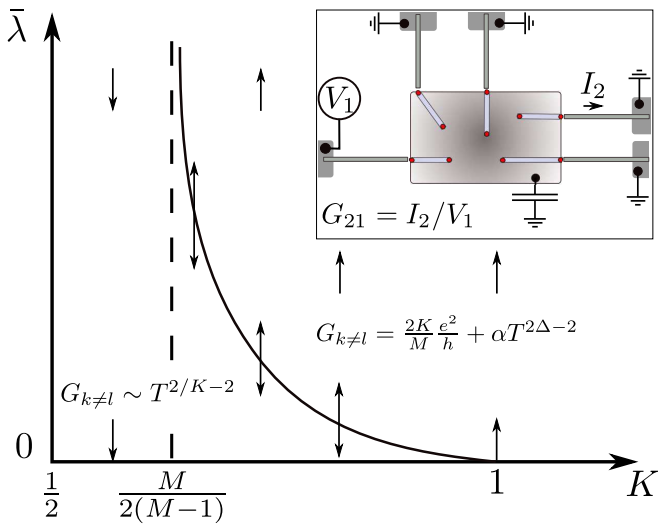}
\caption{The phase diagram 
of the topological Kondo effect in terms of the typical bare $\lambda^+$ coupling $\bar \lambda$ and the Luttinger parameter $K$. The arrows indicate the flow of the couplings under the RG, and $G_{k\neq l}$ is the low temperature Kubo conductance. The curved line separates insulating ($\lambda^+\!\!\!\rightarrow\! 0$) and Kondo ($\lambda^+ \!\!\!\rightarrow\! \infty$) phases. 
The inset shows the sketch of the setup for $M=5$. The central rectangle is an s-wave superconductor and the light bars between the Majoranas are semiconductor nanowires.}
\label{fig:phasediag}
\end{figure}

In what follows, we apply the RG
to our full problem, focusing on $K\leq 1$ and allowing for the anisotropy ($\lambda^+_{jk}\neq\lambda^+$)  and $\lambda^-_{jj}$ terms  arising in a physical realization. The key aspects of the physics will be shown to come from the $\lambda^+_{jk}$ term, extending the relation to QBM to this more general case. 
The results will be used to obtain the low temperature behavior of the Kubo conductance $G_{kl}$ between leads $k,l$. The setup (for $M=5$) is sketched in Fig.~\ref{fig:phasediag} . 

The weak $\lambda_{jk}^{\pm}$ flow under rescaling $a\rightarrow ae^l$ is obtained from the operator product expansion,
\begin{equation}
\frac{d\lambda_{jk}^{+}}{dl}=(1-K^{-1})\lambda_{jk}^{+}+2\nu\sum_{m\neq j,k}\lambda_{jm}^{+}\lambda_{mk}^{+}.
\label{eq:weakRG}\end{equation}
Here $\nu$ is the density of states of the leads. The couplings $\lambda^-_{jj}$ do not renormalize. As the $\lambda^+_{jk}$ terms transfer charge between the leads [see Eq.~\eqref{eq:heffgam}], the first term in Eq.~\eqref{eq:weakRG} promotes a suppression, the second term an enhancement 
of tunneling processes.  This corresponds to the competition of two distinct mechanisms: the suppression of electron tunneling in Luttinger liquids\cite{KFluttlett,*KFluttPRB}, and the enhancement of the coupling due to the Kondo effect. 
Depending on which of these wins one will find a markedly different low temperature behavior, described by the decoupled lead and the Kondo fixed points.

The transition between the two cases is governed by a repulsive, isotropic ($\lambda^+_{jk}=\lambda^+$) intermediate fixed point at $\nu\lambda^{+\star}=\frac{1-K}{2K(M-2)}$, with exponent $\frac{1}{K}-1$ in the relevant, isotropic direction. (The fixed point is attractive from the orthogonal directions.) This is consistent with the weak coupling regime for $K\lesssim 1$, but we believe that the existence of an isotropic fixed point governing the transition remains true in general. 

For $\bar \lambda<\lambda^{+\star}$, where $\bar \lambda$ is the typical bare value of $\lambda^+_{jk}$, 
the low temperature behavior is dictated by the decoupled fixed point. 
In the opposite case, $\lambda^+_{jk}$ flow to large values while becoming more and more isotropic (consistently with the flow around $\lambda^{+\star}$). The crossover to strong coupling is at the scale of the Kondo temperature, $T_{\rm K}\sim E_c e^{-1/\nu\bar\lambda}$. 
The analysis of the strong coupling regime becomes transparent after rotating  $\boldsymbol{\theta}=(\theta_1,\ldots,\theta_M)$, $\boldsymbol{\varphi}=(\varphi_1,\ldots,\varphi_M)$ to decompose them to  $R_0=\mathbf{v}_0\cdot \boldsymbol{\theta}$, $K_0=\mathbf{v}_0\cdot \boldsymbol{\varphi}$ [with $\mathbf{v}_0=\frac{1}{\sqrt M}(1,\ldots,1)$] and to $M-1$ components $r_j$, $k_j$ along the directions orthogonal to $\mathbf{v}_0$. 
This canonically decouples the $K_0,R_0$ sector, while the $\mathbf{r},\mathbf{k}$ sector has $\medop{\sum_j} H_0(k_j,r_j)$ perturbed by
\begin{equation}
H_{\text{eff}}=-\!\!\medop\sum_{j\neq k}\lambda_{jk}^{+}\frac{e^{i(\mathbf{w}_{k}-\mathbf{w}_{j})\cdot\mathbf{r}/2}}{a}-\medop\sum_{j}\frac{\lambda_{jj}^{-}}{2\pi}\mathbf{w}_{j}\cdot\partial_{x}\mathbf{k},
\label{eq:qbmform}
\end{equation}
where $\mathbf{w}_{j}\cdot\mathbf{w}_{l}=\delta_{jl}-\frac{1}{M}$, implying that the minima of the first term form a (hyper)triangular lattice.
Near the Kondo fixed point $\lambda^+\rightarrow \infty$, $\mathbf{r}$ tends to be pinned to one of these minima. This, at the same time,  suppresses\cite{Nayak99,*Oshikawa06}  $\partial_x\mathbf{k}$, thus the $\lambda^-_{jj}$ term can be neglected in this regime. The $\lambda^-_{jj}$ couplings, therefore, only give small, marginal perturbations (through the $R_0$, $K_0$ sector) again, showing that the QBM model indeed captures the essential physics in both the weak and strong coupling regimes. 

Using QBM results\cite{YiKane} we immediately find that the leading perturbations at the Kondo fixed point, given by $\mathbf{r}$ tunneling  between the adjacent minima, have dimension $\Delta=\frac{2K(M-1)}{M}$. The Kondo fixed point is thus stable as long as $K>\frac{M}{2(M-1)}>\frac{1}{2}$; 
it is robust against weak conduction electron interactions and its robustness is enhanced as $M$ increases. 
Due to their non-integer dimension, these charge conserving processes do not admit a free fermion description. 
The SO($M$) Kondo problems thus give rise to NFL behavior 
even with Fermi liquid ($K=1$) leads. 
Such stable NFL fixed points are absent from conventional Kondo systems\cite{NozBlan,*Mat95,*CoxZaw,*Oreg03,*potok2007observation}. Our method allows us to prove that this remarkable feature is a generic property of the topological Kondo effect, vastly generalizing the $M=3,K=1$ result of Ref.~\onlinecite{BeriTK}.

In particular, for $M=4,K=1$ we find
$\Delta=3/2$, as for the two-channel Kondo fixed point\cite{affleck1990current,*affleck1991kondo,*affleck1991critical}. 
Through relabeling $\psi_j$ one can indeed map our problem to the two-channel Kondo model as $\lambda^+_{jk}\rightarrow \lambda^+$, $\lambda^-_{jj}\rightarrow 0$. The topological Kondo effect thus provides a long sought-after stable realization of this fixed point that does not hinge on (but is robust against) having NFL leads, in contrast to earlier proposals\cite{FabGog2c,*FieteNayak,*Law2c,*TsvelikIsing}.  
Placing the two channel Kondo in the QBM context, we also find a new theoretical 
perspective, alternative to Refs.~\onlinecite{affleck1990current,*affleck1991kondo,*affleck1991critical,EK,ColemanKondo1,*ColemanKondo2,*SchoKondo}.

We can now apply our findings to 
the Kubo conductance 
$G_{kl}$. 
The phase diagram in terms of $\bar \lambda$ 
and $K$ is sketched in Fig.~\ref{fig:phasediag}.
Its topology is dictated by the QBM\cite{YiKane}, but the physical meaning of the phases is specific to the topological Kondo effect. 
Tuning $\bar \lambda$ or $K$, the system undergoes a quantum phase transition, switching  $G_{k\neq l}$ from $0$ to 
$\frac{2K}{M}\frac{e^2}{h}$ at $T\rightarrow 0$. 
The transition using $\bar \lambda$ is especially appealing, as $\bar \lambda$ is gate tunable, in principle, in the nanowire realizations of recent experiments\cite{Mourik25052012}. 

Near the decoupled lead fixed point, $G_{kl}$ vanishes as $G_{kl}\sim T^{2/K-2}$ as $T\rightarrow 0$. This is the known suppression of electron tunneling between Luttinger liquid leads\cite{KFluttlett,*KFluttPRB}, with the exponent coming from the first term of \eqref{eq:weakRG}.

For $T\ll T_{\text K}$ near the Kondo fixed point,  we have $G_{kl}=\frac{2Ke^2}{h}(\delta_{kl}-\frac{1}{M})+\alpha_{kl}T^{2\Delta-2}$, where $\alpha_{kl}$ are nonuniversal constants. The fixed point ($T\rightarrow 0$) value follows from the emergent boundary conditions $\mathbf{r}=$ fixed, $\partial_x\mathbf{k}=0$ at $\lambda^+\rightarrow\infty$, and can be obtained from an immediate generalization of the calculations of Ref.~\onlinecite{Nayak99,*Oshikawa06}. Note that $G_{ll}=\frac{e^2}{h}\Delta$. A stable Kondo fixed point ($\Delta>1$) thus comes with $G_{ll}$ violating the $\frac{e^2}{h}$ limit for single-channel normal conduction. This is due to the emergent boundary conditions translating into correlated, multiparticle Andreev processes by which holes, not only electrons can be backscattered\cite{Nayak99,*Oshikawa06}. 
The $T^{2\Delta-2}$ dependence is due to second order corrections in processes tunneling $\mathbf{r}$ between the minima in Eq.~\eqref{eq:qbmform}. (The first order corrections of the current-current correlators underlying $G_{kl}$ vanish\cite{2CKnote}.) The convergence to an enhanced conductance through such nontrivial power laws gives a clear signature of the NFL Kondo physics.

In conclusion, we have introduced a bosonization based picture for Majorana-lead couplings relevant to ongoing transport experiments. The key feature is the appearance of Klein factors $\Gamma_j$, virtually extending the number of Majoranas in the system. We have illustrated the utility of our approach applying it to two-terminal normal-superconductor transport with interacting leads and solving the SO($M$) topological Kondo problem. We expect that our picture will be a useful starting point for a number of new 
problems exploring the interplay of Majorana fermions and strong correlations.  

I thank N. d'Ambrumenil, F. Hassler,  D. Schuricht for useful discussions and especially N. R. Cooper for his advice and a previous collaboration\cite{BeriTK}. This work was supported by a MC IEF Fellowship.

\ifx\mcitethebibliography\mciteundefinedmacro
\PackageError{apsrevM.bst}{mciteplus.sty has not been loaded}
{This bibstyle requires the use of the mciteplus package.}\fi


\begin{mcitethebibliography}{42}
\bibitem[{M4f()}]{M4fn}
\bibinfo{note}{Here we assumed that all $\gamma_j\gamma_k$ are different which
  holds unless $M_\text{tot}=M=4$; our results, however, extend to this case
  too due to the reducibility of spinors for $M_\text{tot}>M=4.$}\relax
\mciteBstWouldAddEndPuncttrue
\mciteSetBstMidEndSepPunct{\mcitedefaultmidpunct}
{\mcitedefaultendpunct}{\mcitedefaultseppunct}\relax
\EndOfBibitem
\bibitem[{2CK()}]{2CKnote}
\bibinfo{note}{For $M=4$ $K=1$ this underlies the difference to the usual
  $\sqrt{T}$ two-channel Kondo result showing that $G_{kl}$ does not map to the
  conductance in the two-channel Kondo language.}\relax
\mciteBstWouldAddEndPunctfalse
\mciteSetBstMidEndSepPunct{\mcitedefaultmidpunct}
{}{\mcitedefaultseppunct}\relax
\EndOfBibitem
\end{mcitethebibliography}
\end{document}